\documentclass[epj]{svjour}

\usepackage{amssymb,latexsym}
\usepackage{graphicx}
\usepackage{fancyhdr}
\def\makeheadbox{{
 }}
\def\mytitle{My title} 
\def\myauthors{My name}  
\def\mytype{My type of session}
\def\mysession{My session}

\def\mytitle{1-loop Corrections to the $\rho$ parameter
 in the Left-Right Twin Higgs Model} 
\def\myauthors{Dong-Won Jung}    
\def\mytype{Contributed Talk}    
\def\mysession{Alternatives}

\begin{document}
\title{1-loop Corrections to the $\rho$ Parameter in the
 Left-Right Twin Higgs Model}
\author{Dong-Won Jung
\thanks{\emph{Email:} dwjung@ncu.edu.tw, now moved to Nat'l Central Univ., Taiwan, after Sep. 1.}%
\and
 Jae Yong Lee
}                     
%
%
\institute{Korea Institute for Advanced Study, Hoegiro 87 Dongdaemun-gu, Seoul 130-722, Korea }
%
\date{}
\abstract{
We implement a one-loop analysis of the $\rho$ parameter in the
Left Right Twin Higgs model, including the logarithmically
enhanced contributions from both fermion and scalar loops.
Numerical results show that the one-loop contributions are
dominant over the tree level corrections in most regions of
parameter space. The experimentally allowed values of
$\rho$-parameter divide the allowed parameter space into two
regions; less than $670~{\rm GeV}$ and larger than $1100~{\rm
GeV}$ roughly, for symmetry breaking scale $f$. Our numerical
analysis significantly reduces the parameter space which are
favorably accessible to the LHC.
\PACS{
      {12.60.-i}{Models beyond the standard model}   \and
      {12.15.Lk}{Electroweak radiative corrections}
     } 
} 
\maketitle
\section{Introduction}
\label{intro}
The Standard Model (SM) has excellently described high energy
physics up to energies of ${\mathcal O}(100)$ GeV. The only
undetected constituent of the SM, up to now, is a Higgs boson
which is required to explain the generation of fermion and gauge
boson masses. Theoretically, the Higgs boson mass squared is
quadratically sensitive to any new physics scale beyond the
Standard Model (BSM) which may arise at higher energy scales and
hence stabilization of the Higgs mass squared prefers the energy
scale at which the BSM turns up to be lowered to ${\mathcal O}(1)$
TeV. On the other hand, electroweak precision measurements with
naive naturalness assumption raise the energy scale of the BSM up
to 100 TeV or even higher. Hence, there remains a tension between
theory and experiment associated with the stabilization of the SM
Higgs mass. But with the start-up of the LHC the tension
may be relaxed by direct observation of the BSM at TeV energy scale.
 The idea of little Higgs originates in the speculation that the SM
Higgs may be a pseudo-Nambu-Goldstone
boson~
\cite{Georgi:1974yw,Kaplan:1983fs,Arkani-Hamed:2001nc,Arkani-Hamed:2002pa,Arkani-Hamed:2002qx,Arkani-Hamed:2002qy,Schmaltz:2002wx}.
Stabilization of the Higgs mass in the little Higgs theories 
is achieved by the "collective
symmetry breaking" which naturally renders the SM Higgs mass much
smaller than the symmetry breaking scale. The distinct elements of
little Higgs models are a vector-like heavy top quark and various
scalar and vector bosons. The former is universal while the latter
is model-dependent. Both of them contribute significantly to
one-loop processes and hence establish strict constraints on the
parameter space of little Higgs models. At worst, electroweak
precision tests push up the symmetry breaking scale to $5$ TeV or
higher, and regenerate significant fine-tuning in the Higgs
potential.
 Twin Higgs idea shares the same origin with that of little
Higgs in that the SM Higgs is a pseudo-Nambu-Goldstone
boson~\cite{Chacko:2005pe}. But rather than using collective
symmetry breaking to stabilize the Higgs mass squared it makes
 use of additional discrete symmetry. In
other words, the discrete symmetry ensures the absence of
quadratic divergence in the Higgs mass squared. 
The twin Higgs mechanism is realized by identifying the
discrete symmetry with left-right symmetry in the left-right
model~\cite{Chacko:2005un}. The left-right twin Higgs (LRTH) model
contains $U(4)_1\times U(4)_2$ global symmetry as well as
$SU(2)_L\times SU(2)_R\times U(1)_{B-L}$ gauge symmetry. The
left-right symmetry acts on only the two $SU(2)$'s gauge symmetry.
A pair of vector-like heavy top quarks play a key role at
triggering electroweak symmetry breaking just as that of the
little Higgs theories. Besides, the other Higgs particles acquire
large masses not only at quantum level but also at tree level.
These heavy Higgs bosons make the model deliver rich
phenomenology at the LHC~\cite{Goh:2006wj}.  
But theoretically, they lead to large
radiative corrections to one-loop processes and, in return, the
allowed parameter space can be reduced significantly. In this
paper, we perform a one-loop analysis of the $\rho$-parameter in
the LRTH model
 to reduce the parameter space. This is based on the original 
work with Jae Yong Lee, KIAS~\cite{Jung:2007ea}.

\section{Left-Right Twin Higgs Model~\cite{Goh:2006wj}}
\label{sec:1} 
 The LRTH model is based on the
global $U(4)_1\times U(4)_2$ symmetry, with a locally gauged
subgroup $SU(2)_L\times SU(2)_R\times U(1)_{B-L}$. A pair of Higgs
fields, $H$ and $\hat H$, are introduced and each transforms as
$(4,1)$ and $(1,4)$ respectively under the global symmetry. They
are written as
\begin{equation}
H=\left(\begin{array}{c} H_L\\ H_R\end{array}\right),\qquad
\hat H=\left(\begin{array}{c} \hat H_L\\ \hat H_R\end{array}\right),
\end{equation}
where $H_{L,R}$ and $\hat H_{L,R}$ are two component objects which are charged under
the $SU(2)_L\times SU(2)_R\times U(1)_{B-L}$ as
\begin{equation}
H_L\mbox{ and }\hat H_L\,:\,(2,1,1),\qquad
H_R\mbox{ and }\hat H_R\,:\,(1,2,1).
\end{equation}
The global $U(4)_1\,(U(4)_2)$ symmetry is spontaneously broken down to its subgroup
$U(3)_1\,(U(3)_2)$ with VEVs
\begin{equation}
\langle H\rangle =\left(\begin{array}{c} 0\\ 0 \\ 0 \\ f\end{array}\right),\qquad
\langle \hat H \rangle = \left(\begin{array}{c} 0\\ 0\\ 0\\ \hat f \end{array}\right).
\end{equation}
Each spontaneous symmetry breaking results in seven Nambu-Goldstone bosons,
which are parameterized as
\begin{equation}
H=fe^{\pi/f}\left(\begin{array}{c} 0 \\ 0\\ 0\\ 1\end{array}\right),~
\pi=\left(\begin{array}{cccc} -\frac{N}{2\sqrt{3}} & 0 & 0 & h_1 \\
0 & -\frac{N}{2\sqrt{3}} & 0 & h_2 \\
0 & 0 & -\frac{N}{2\sqrt{3}} & C \\
h^\ast_1 & h^\ast_2 & C^\ast & \frac{\sqrt{3}N}{2}\end{array}\right),
\end{equation}
where $\pi$ is the corresponding Goldstone field matrix. $N$ is a neutral real field
$C$ and $C^\ast$ are a pair of charged complex scalar fields, and $h_{SM}=(h_1,h_2)$
is the SM $SU(2)_L$ Higgs doublet.
$\hat H$ is parameterized in the identical way by its own Goldstone boson matrix,
$\hat \pi$, which contains $\hat N$, $\hat C$, and $\hat h=(\hat h^+_1,\hat h^0_2)$.
 In turn, two $U(4)/U(3)$'s symmetry breaking left with fourteen Nambu-Goldstone bosons.
The linear combination of $C$ and $\hat C$, and the linear combination
of $N$ and $\hat N$ are eaten by the gauge bosons of $SU(2)_R\times U(1)_{B-L}$,
which is broken down to the $U(1)_Y$.
The orthogonal linear combinations, a charged complex scalar $\phi^\pm$
and a neutral real pseudoscalar $\phi^0$, remain as Nambu-Goldstone bosons.
On top of that, the SM Higgs acquires a VEV, $\langle h_{SM}\rangle=(0,v/\sqrt{2})$,
and thereby electroweak symmetry
$SU(2)_L\times U(1)_Y$ is broken down to $U(1)_{EM}$. But $\hat h$'s
do not get a VEV and remain as Nambu Goldstone bosons. These Nambu Goldstone 
bosons acquire masses through quantum effects and/or soft symmetry breaking 
terms, so called $\mu$-terms,
\begin{equation}
V_{\mu}=-\mu^2_r(H^\dagger_R\hat H_R+c.c.)+\hat \mu^2 \hat H^\dagger_L \hat H_L,
\end{equation}
which contribute to the Higgs masses at tree level.
Because of the extend gaugue symmetry, there are extra gauge bosons besides the
SM gauge bosons, $W_H$ and $Z_H$, masses of which are proportional to $f$ and
 $\hat f$. The existence of the extra gauge bosons would be the typical feature
 of the generic left-right symmetric models.

To cancel the quadratic sensitivity of the Higgs mass to the top quark loops,
a pair of vector-like, charge $2/3$ fermion $({\mathcal Q}_L,{\mathcal Q}_R)$
are incorporated into the top Yukawa sector,
\begin{eqnarray}
&&{\mathcal L}_{Yuk}=\\ \nonumber 
&&y_L\bar Q_{L3}\tau_2 H^\ast_L {\mathcal Q}_R+
y_R\bar Q_{R3}\tau_2 H^\ast_R {\mathcal Q}_L-M\bar{\mathcal Q}_L {\mathcal Q_R}+h.c.,
\end{eqnarray}
where $\tau_2=\left(\begin{array}{cc} 0 & -1 \\ 1 & 0
\end{array}\right)$, $Q_{L3}=-i(u_{L3},d_{L3})$ and
$Q_{R3}=(u_{R3},d_{R3})$ are the third generation up- and
down-type quarks, respectively. The left-right parity indicates
$y_L=y_R(\equiv y)$. The mass parameter $M$ is essential to the
top mixing. The value of $M$ is constrained by the $Z\to b\bar b$
branching ratio. It can be also constrained by the oblique
parameters, which we will do in the letter. Furthermore, it yields
large log divergence of the SM Higgs mass. To compensate for it
the heavy gauge bosons also get large masses by increasing the
value of $\hat f$. Therefore it is natural for us to take
$M \lesssim yf$.
\section{Results and Discustion}
\begin{figure}
\includegraphics[width=0.45\textwidth,height=0.36\textwidth,angle=0]{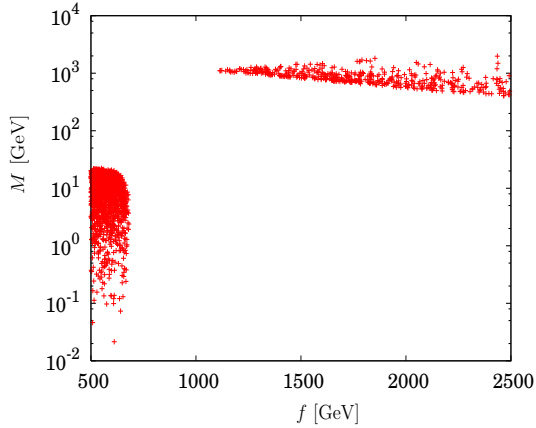}
\caption{$M ~{\bf vs.} f$, allowed range.}
\label{fig:fm}       
\end{figure}

\begin{figure}
\includegraphics[width=0.45\textwidth,height=0.36\textwidth,angle=0]{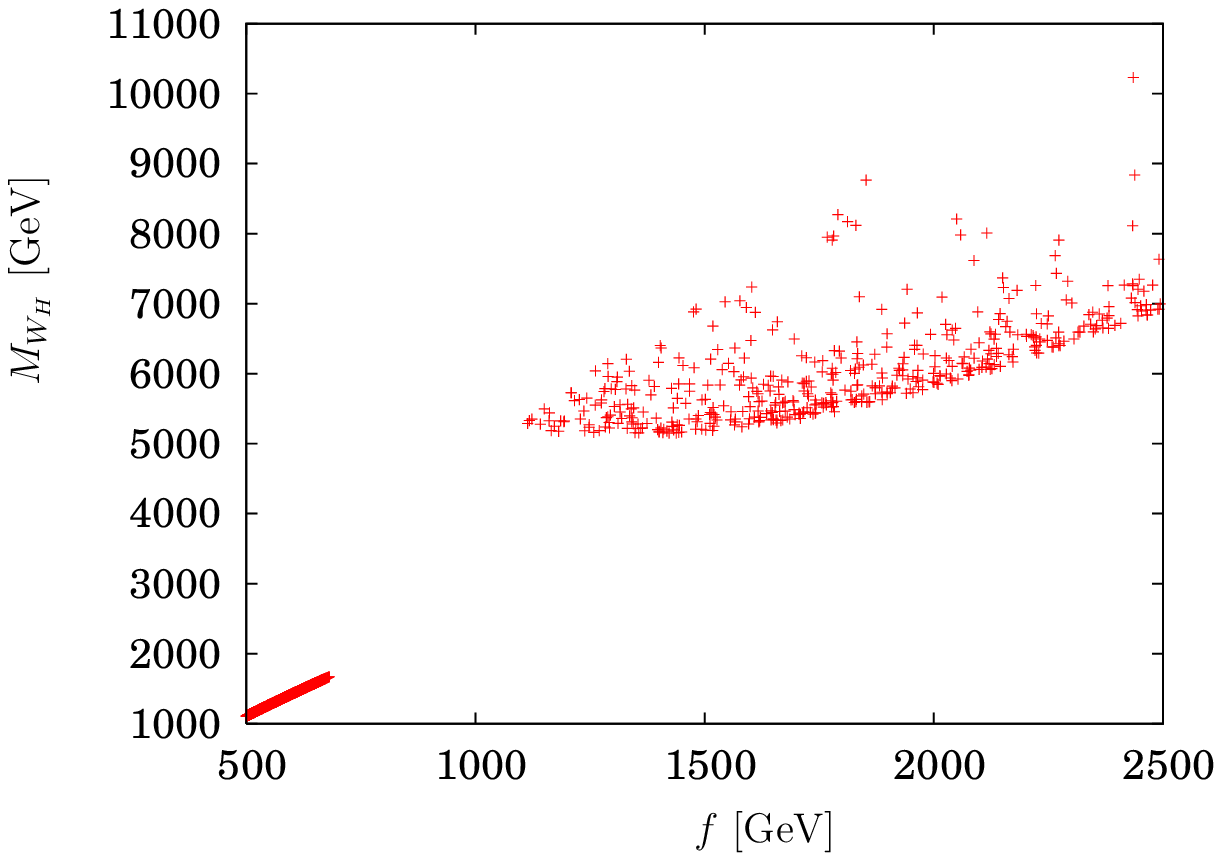}
\caption{$m_{W_H} ~{\bf vs.} f$, allowed range.}
\label{fig:fw}       
\end{figure}

\begin{figure}
\includegraphics[width=0.45\textwidth,height=0.36\textwidth,angle=0]{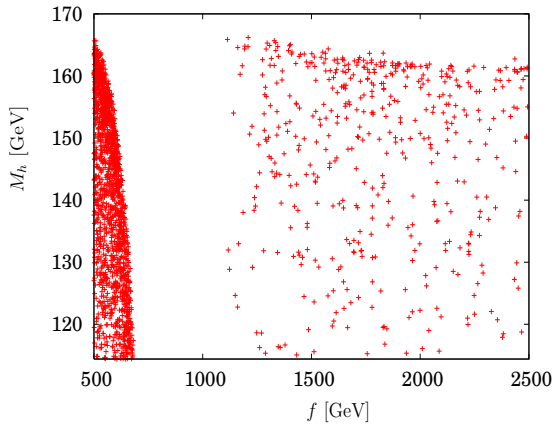}
\caption{$m_h ~{\bf vs.} f$, allowed range.}
\label{fig:fh}       
\end{figure}

The $Z$-pole, $W$-mass,
and neutral current data can be used to search for and set limits
on deviations from the SM. In the article we concentrate
particularly on the the $\rho$-parameter, which is defined as
\begin{equation}\label{eq:defrho}
\rho\equiv\frac{M^2_W}{M^2_Zc^2_\theta}.
\end{equation}
The effective leptonic mixing angle $s^2_\theta(=1-c^2_\theta)$
at the $Z$-resonance is defined
as the ratio of the electron vector to axial vector coupling constants to the $Z$-boson,
\begin{equation}
\frac{Re(g^e_V)}{Re(g^e_A)}\equiv4s^2_\theta-1,
\end{equation}
where the coupling constants of a fermion $\psi$ to the gauge boson $X$ is given as,
\begin{equation}
{\mathcal L}=i \bar \psi_1\gamma_\mu(g_V+g_A\gamma_5)\psi_2 X^\mu.
\end{equation}
Using the procedure in Ref.~\cite{Chen:2003fm}, we can calculate the 1-loop
 corrected W boson mass 
\begin{equation}\label{eq:mw}
M^2_W=\frac{1}{2}\Big[a(1+\Delta \hat r)+\sqrt{a^2(1+\Delta \hat r)^2+4a\Pi^{WW}(0)}\Big],
\end{equation}
with $ \quad a\equiv\frac{\pi\alpha(M_Z)}{\sqrt{2}G_Fs^2_\theta}$, and the 
definition of $\Delta \hat r$ is
\begin{eqnarray}
\Delta {\hat r}=
&&-\frac{\Delta s^2_\theta}{s^2_\theta}-\frac{Re(\Pi^{ZZ}(M^2_Z))}{M^2_Z}
+\Pi^{\gamma\gamma'}(0) \\ \nonumber 
&&+2(\frac{g^e_V-g^e_A}{Q_e})\frac{\Pi^{\gamma Z}(0)}{M^2_Z}
-\frac{c^2_\theta-s^2_\theta}{c_\theta s_\theta}
\frac{Re(\Pi^{\gamma Z}(M^2_Z))}{M^2_Z}. 
\end{eqnarray}
The 1-loop corrected 
$\rho$ parameter is then obtained using Eq.~(\ref{eq:defrho})
with the $M^2_W$ value predicted in Eq.~(\ref{eq:mw}).
 For doing the calculation concerning the precision measurements,
the standard experimental values are necessary which play as input
parameters. Here, we use the following experimentally measured
values for the input parameters~\cite{Yao:2006px,:2005em}:
\begin{eqnarray}
G_F&=& 1.16637(1)\times 10^{-5}\mbox{ GeV}^{-2},\\
M_Z&=& 91.1876(21)\mbox{ GeV},\\
\alpha(M_Z)^{-1}&=&127.918(18),\\
s^2_\theta&=&0.23153(16).
\end{eqnarray}
We also take the top and bottom quark masses as
~\cite{Yao:2006px,Rodrigo:1997gy}
\begin{equation}
m_t = 172.3 ~{\rm GeV}, \qquad m_b = 3 ~{\rm GeV},
\end{equation}
where $m_t$ is the central value of the electroweak fit and $m_b$
is the running mass at the $M_Z$ scale with $\overline{MS}$ scheme.
Including all the SM corrections (top quark loop, bosonic loops),
 we take the allowed range of $\rho$ parameter as~\cite{Yao:2006px}
\begin{equation}
1.00989 ~\leq ~\rho^{exp}~\leq ~1.01026.
\end{equation}
The input parameters of the LRTH model~\cite{Chacko:2005un} are as
follows:
\begin{equation}
f, ~M, ~\mu_r, ~\hat \mu,
\end{equation}
where $M$ is the heavy top quark mass scale, both $\mu_r$ and
$\hat \mu$ are soft symmetry breaking terms. The masses of the top
and heavy top quarks are determined by $f$ and $M$ while those of
the scalar particles $\hat h_1,~\hat h_2,~\phi^\pm$ and $\phi^0$
largely depend on $\hat \mu,~\mu_r$ and $f$. Another scale $\hat
f$, which is associated with the masses of the heavy gauge bosons,
can be determined from the electroweak symmetry breaking
condition: there is a generic relation between $\hat f$ and $f$
since Coleman-Weinberg potential of the Higgs boson mostly depends
on $M, f$ and $\hat f$. For scalar potential, there is a tree
level mass term proportional to $\mu_r^2$. So we may not acquire
negative mass squared term which is necessary for electroweak
symmetry breaking and it gives an upper bound for the value of
$\mu_r$.
For a given $f$, $\hat f$ becomes larger as $M$
increases. It is because the increase of $M$ contributes
positively to the Higgs mass through the top loop while the
increase of $\hat f$ contributes negatively to the Higgs mass
through the gauge boson loop, and thereby these contributions
cancel out themselves in order to retain $v=246$ GeV. 
To draw a meaningful information on the model parameters from the
$\rho$-parameter , we scan the parameter space generally, i.e.,
\begin{equation}
500 ~{\rm GeV} ~\leq ~f ~\leq ~2500 ~{\rm GeV},~~~ 0~ \leq ~M,
~\mu_r, ~\hat \mu ~\leq ~f.
\end{equation}
Even though too large $f$ makes the model unviable, we take the
rather large value of $f$, 2.5  TeV, as an upper limit for
completeness of the scanning. As a result of $\rho$-parameter
calculation, we can obtain the allowed regions of parameter space.
As an example, Fig.~\ref{fig:fm}  shows the allowed regions of
parameter space for $f$ versus $M$. 
It is interesting to notice that the allowed parameter
space is divided into two regions; less than 670 GeV and larger
than 1100 GeV roughly, for $f$. This can be figured out as
follows. The loop corrections tend to be larger as $f$ increases.
It is because the masses of the particles involved in one-loop
correction increase in general as $f$ increases. But at the same
time, the mixing angles of top-heavy top quarks also vary. Since
the mixing angles depend on not only $f$ but also $M$, these two
effects compete during the increase of $f$. Because of this
interplay of top mixing angles and masses, we have two distinct
allowed parameter spaces. For small $f$, solution points prefer
very small values of $M$. It means there is no large mixing
between the top and heavy top quarks. In general, $\Pi^{WW}(0)$ is
large for small $f$, and
 decreases as $f$ increases. So for fitting the observed W-boson
 mass in the small $f$ region, which is directly related to the
$\rho$-parameter, we restrict the $\Delta \hat r$ within rather
small range. Because the $\Delta \hat r$ is mostly determined by
$\Pi^{ZZ}(M_Z^2)$, it should be also small. For doing that, we
should take the small value of $M$, which makes the masses and
mixing angles of heavy top quark small. We find that in the small
$f$ region, $M$ should be smaller than about $22~{\rm GeV}$.
 Soft symmetry breaking parameter $\mu_r$ is restricted to the
 values less than around $60~{\rm GeV}$.
 This bound arises mainly from the electroweak symmetry breaking
condition, and is generically independent of the $\rho$-parameter.
Another free parameters $\hat \mu$ is not restricted from the
one-loop corrected $\rho$-parameter. The reason is that $\hat \mu$
only contributes to the masses of $\hat h_1$ and $\hat h_2$, and
their contributions are effectively cancelled among the relevant
loop diagrams. 
This region of parameter space can provide constraints on the
masses of many particles appear in this model. First, let us
consider the masses of the heavy top and heavy gauge bosons. 
Their masses generically increase
as $f$ increases. The mass of the heavy top quark is uniquely
determined when $f, \hat f$ and $M$ are fixed. So does top Yukawa
coupling. Basically $\hat f$ is determined by the electroweak
symmetry breaking condition, but  their $M$ and $\mu_r$ dependence
provoke the ambiguity on its value. For small $f$ region, since
$M$ is also very small, the $m_T$ is almost determined by $f$
alone.  For large $f$ region, it becomes spread due to the top mixing
angles. 
The plots of the heavy $Z$ and $W$ boson masses versus $f$
are quite similar to that of the heavy top mass versus $f$. In the
case of heavy $W$ boson, the strongest constraint come from $K_L -
K_S$ mixing. The strongest bound ever known is  $m_{W_H} > 1.6$
TeV, with the assumption of $g_L = g_R$~\cite{Beall:1981ze}. This
can exclude some region from Fig.~\ref{fig:fw}. In this
case, small $f$ region can be completely excluded. 
If the lower bound for $f$ is confirmed, 
we can give the lower bound for $f$ as $1.1 ~{\rm
TeV}$ from our calculation of the $\rho$-parameter and also for
many particles appear in the model. Another constraints on the
$m_{W_H}$ from CDF and D0 are about $650 \sim 786 {\rm GeV}$, as
lower bound~\cite{Affolder:2001gr,Abachi:1995yi}. For Our results
remain safe from these experimental bounds. Heavy $Z$ boson has
also been studied in detail by many experimentalists. Current
experimental bound is about $500 \sim 800 {\rm GeV}$ from
precision measurements~\cite{Yao:2006px} and $\sim 630 {\rm GeV}$
from CDF~\cite{Yao:2006px}. In this case, also safe is the mass of
heavy $Z$ boson.
With the parameters allowed by the $\rho$-parameter, the masses of
new scalar bosons $\hat h_{1,2}, \phi^0$ and $ \phi^\pm$ are also
constrained. $\hat h_{1,2}$ has almost degenerate masses, and are
dependent on both $\mu_r$ and $\hat \mu$, unlike the
$\phi^{0,\pm}$ which depend only on $\mu_r$.
 Their masses are seriously constrained according to the value of $f$.
 Unfortunately, we cannot give a lower bound on the mass of $\phi^0$.
 In fact, its mass, though it is quite small, arise from radiative corrections.
 For $\phi^\pm$, the loop
contribution is rather large so it acquire larger mass compared to
the neutral one. 
The $\rho$-parameter cannot give a strong restriction on the Higgs
mass. In the whole space, Higgs mass is restricted below about
$167~{\rm GeV}$. We cannot give a lower bound for Higgs boson mass
from $\rho$-parameter itself. Here, we adopt the LEP bound for
Higgs mass, $114.4~{\rm GeV}$~\cite{Barate:2003sz}, since its
structure is same as the SM. The generic behavior of Higgs mass as
a function of $f$ is shown in Fig~.\ref{fig:fh}.

%
%
\section{Summary}
We summarize the results of our analysis as follows. With the
observed $\rho$-parameter, we see that the allowed parameter space
is divided into two separate regions: $f$ smaller than about  670 GeV and larger than about
1.1 TeV. We give the bounds on the mass spectrum of many particles
for either region. Especially the heavy gauge bosons remain safe
from the experimental constraints. Unlike the other particles, we
cannot set a lower bound for the neutral $\phi^0$ scalar. But loop correction plays
an important role for the charged $\phi^\pm$ scalars, yielding mass difference
between the charged and neutral scalars. Further analysis is
required in order to reduce the allowed region. If the small $f$
region is excluded, for example by Ref.~\cite{Goh:2006wj},
we can provide exact lower bounds for the
masses of $T, Z_H, W_H, \hat h_{1,2}$, and $ \phi^\pm$. But even
in that case, we cannot do so for $\phi^0$ and SM Higgs boson.
%
%

\end{document}